# ensemblQueryR: fast, flexible and high-throughput querying of Ensembl LD API endpoints in R

## Authors

Fairbrother-Browne A[1,3,4], García-Ruiz S[1,2], Reynolds R H[1,2], Ryten M[1,2,*], Hodgkinson A[3,*]

1. Department of Genetics and Genomic Medicine Research & Teaching, UCL GOS Institute of Child Health, London, UK
2. NIHR Great Ormond Street Hospital Biomedical Research Centre, University College London, London, UK
3. Department of Medical and Molecular Genetics, School of Basic and Medical Biosciences, King's College London, London, UK
4. Department of Neurodegenerative Disease, Queen Square Institute of Neurology, UCL, London, UK

*Contributed equally.

## Abstract

We present ensemblQueryR, a package providing an R interface to the Ensembl REST API that facilitates flexible, fast, user-friendly and R workflow integrable querying of Ensembl REST API linkage disequilibrium (LD) endpoints, optimised for high-throughput querying. ensemblQueryR achieves this through functions that are intuitive and amenable to custom code integration, use of familiar R object types as inputs and outputs, code optimisation and optional parallelisation functionality. For each LD endpoint, ensemblQueryR provides two functions, permitting both single-query and multi-query modes of operation. The multi-query functions are optimised for large query sizes and provide optional parallelisation to leverage available computational resources and minimise processing time. We demonstrate that ensemblQueryR has improved performance in terms of random access memory (RAM) usage and speed, delivering a 10-fold speed increase over analogous software whilst using a third of the RAM. Finally, ensemblQueryR is near-agnostic to operating system and computational architecture through availability of Docker and singularity images, making this tool widely accessible to the scientific community.

## Statement of need

### Background

Linkage disequilibrium (LD) is the non-random association of alleles arising from different loci [1]. In population genetics, LD is a measure of the frequency with which an allele of one variant is correlated with an allele of a proximal variant within a particular population [2]. There are many applications for LD measures in genomics workflows. For example, in the context of genome-wide association studies (GWAS), which have been used to detect associations between genetic variants and a wide range of human phenotypes, downstream interrogation of local LD structure is required to identify the potential 'causal' variant at a nominated locus that exerts the effect on the downstream phenotype. Equally, in expression quantitative trait loci (eQTL) analyses, which aim to uncover associations between genetic variants and expression of a *cis* or *trans* gene (eGene), LD information is required for the identification of the potential causal variant affecting expression of the eGene. Further downstream, LD information is useful for functional annotation, where genetic variants or regions in LD with a target variant can illuminate biological processes being affected by the GWAS- or eQTL-implicated target variant. As such, it is important that LD information for a range of human populations can be easily queried by researchers in an efficient and accessible way.

Despite the widespread usage of LD measures in genomic research, the majority of tools available at present are web-based. Although these offer user-friendly interfaces and can be useful for one-off or small queries, they are not ideal for reproducible, R workflow-oriented researchers wishing to submit many large queries. Programmatic tools offer a solution to these problems, however, very few tools for the retrieval of LD metrics exist.

To our knowledge, one R package exists that provides a programmatic interface to LD metric retrieval. LDlinkR (version 1.2.3) [3] provides a programmatic R-based interface to the web-based tool LDlink [4], permitting retrieval of LD metrics using a range of query types. However, LDlinkR has a number of key limitations with respect to speed and query handling. Firstly, the user is required to obtain an access token by signing up on the NCBI website, which is then supplied as an argument to all LDlinkR functions. This requirement is in place to set a limit on user queries, meaning that attempts to speed up the tool using parallelisation will easily exceed query limits and cause the tool to return timeout errors. This can result in the user's access token being blocked. Secondly, a number of functions for retrieving LD metrics are configured for singular queries only – such as the LDpair and LDproxy functions – meaning that the user must write custom code to submit more than one query at one time. As such, although LDlinkR is a useful programmatic alternative to the LDlink webtool, it is not suited to fast, high-throughput multi-query retrieval of LD metrics.

Ensembl is another widely-used source of LD metrics, offering an API which supports an array of query configurations [5,6]. However, some challenges are presented by direct API usage, for example, its usage requires some technical expertise. Additionally, it is not easily integrable with typical R workflows, it precludes input of standard R objects (such as data frames, vectors or lists), it does not output data in an intuitive format and it is not easily adaptable to high-throughput workflows. To our knowledge, no R package has been developed to facilitate querying the Ensembl API and in particular to retrieve Ensembl LD metrics. In light of this, and to address the limitations of current tools, we present ensemblQueryR, an R package that provides fast, efficient, user-friendly querying of Ensembl LD data, with a focus on intuitive, high-throughput, R workflow integration.

ensemblQueryR has been made freely available for use (DOI: 10.5281/zenodo.7837882) [7,8]. The package can also be used in Docker [9] or singularity [10] containers, for which the images can be found on Docker Hub [11] or the singularity image repository [12].

## Implementation

### Our approach

ensemblQueryR provides a suite of functions that wrap around three Ensembl API 'endpoints'. These endpoints operate to retrieve data from Ensembl databases through the following query configurations:

1. Window: retrieval of LD metrics for a variant and all other variants in a window around the target variant;
2. Pair: retrieval of LD metrics between a pair of target variants;
3. Region: retrieval of LD metrics between all pairs of variants in a defined target region.

Single-query and multi-query wrapper functions are provided for each of these Ensembl API endpoints, all of which are described in further detail in **table 1**.

| ensemblQueryR function | Ensembl API endpoint | Arguments | Output | Description |
|---|---|---|---|---|
| ensemblQueryGetPops | Information [5] | N/A | A list of human populations (1000 Genomes, Gambian Genome Variation Project) for which LD metrics can be retrieved. The strings in the 'name' column can be supplied to the 'pop' argument in all ensemblQueryR functions. For further information on the available populations, see Ensembl [19]. | This function retrieves a list of the Ensembl populations for which LD metrics can be queried. These can be supplied to the 'pop' argument across ensemblQueryR functions. |

| Function | Category | Inputs | Output | Description |
|---|---|---|---|---|
| pingEnsembl | Information [5] | N/A | An integer (and message) to indicate the status of the Ensembl API. Returns 1 and reports "Server OK." if the server is up. | This function checks and informs the user of the status of the Ensembl API. |
| ensemblQueryLDwithSNPwindow | Window [5] | rsid<br>r2<br>d.prime<br>window.size<br>pop | A data frame with five columns: 'query' (the variant input to 'rsid'), snp_in_ld (variant(s) in LD with 'query'), r2 (r-squared statistic), d_prime (D' statistic), population_name (the population supplied to 'pop'). | This function retrieves variants in LD with the query variant, within a given genomic window. |
| ensemblQueryLDwithSNPwindowDataframe | Window [5] | in.table<br>r2<br>d.prime<br>window.size<br>pop<br>cores | A data frame with five columns: 'query' (the variant input to 'rsid'), snp_in_ld (variant(s) in LD with 'query'), r2 (r-squared statistic), d_prime (D' statistic), population_name (the population supplied to 'pop'). | This function takes a data frame with a column of variant rsIDs (Reference SNP cluster IDs). It retrieves variants in LD with each query variant within a given genomic window. |
| ensemblQueryLDwithSNPpair | Pair [5] | rsid1<br>rsid2<br>pop | A data frame with five columns: 'query1' (the variant input to 'rsid1'), 'query2' (the variant input to 'rsid2'), r2 (r-squared statistic), d_prime (D' statistic), population_name (the population supplied to 'pop'). | This function takes a pair of rsIDs and retrieves their LD metrics. |
| ensemblQueryLDwithSNPpairDataframe | Pair [5] | in.table<br>pop<br>cores | A data frame with five columns: 'query1' (the variant input to 'rsid1'), 'query2' (the variant input to 'rsid2'), r2 (r-squared statistic), d_prime (D' statistic), population_name (the population supplied to 'pop'). | This function takes a data.frame containing paired rsIDs, retrieving LD metrics (D' and $R^2$) for all pairs. |
| ensemblQueryLDwithSNPregion | Region [5] | chr<br>start<br>end<br>pop | A data frame with eight columns: 'query_chr' (the query chromosome supplied to 'chr'), 'query_start' (the query start coordinate supplied to 'start'), 'query_end' (the query end coordinate supplied to 'end'), 'rsid1' (variant one of two in the pair), 'rsid2' (variant two of two in the pair), r2 (r-squared statistic), d_prime (D' statistic), population_name (the | This function takes a genomic coordinate, retrieving LD metrics (D' and $R^2$) for all rsID within the defined region. |

| | | | population supplied to 'pop'). | |
|---|---|---|---|---|
| ensemblQueryLDwithSNPregionDataframe | Region [5] | in.table<br>pop<br>cores | A data frame with eight columns: 'query_chr' (the query chromosome supplied to 'chr'), 'query_start' (the query start coordinate supplied to 'start'), 'query_end' (the query end coordinate supplied to 'end'), 'rsid1' (variant one of two in the pair), 'rsid2' (variant two of two in the pair), r2 (r-squared statistic), d_prime (D' statistic), population_name (the population supplied to 'pop'). | This function takes a data frame containing genomic coordinate(s) and retrieves LD metrics (D' and $R^2$) for all rsID within the defined region(s). |

**Table 1. Table to describe the functions comprising the ensemblQueryR package and their relationship to the three LD Ensembl API endpoints.**

For ensemblQueryR to be useful in a high-throughput context, the main challenge to be surmounted was that the Ensembl API endpoints are configured to handle single queries. To address this, ensemblQueryR's three multi-query functions (with names ending in 'Dataframe', described in **Table 1**) take data frame objects as input, where each row will be submitted as a separate query to the Ensembl API. The base R lapply function (Version 4.0.5) [13] is then used to apply the corresponding single-query function over the input data frame, iteratively formulating an API query from each data frame row.

Building on ensemblQueryR's high-throughput capabilities, we implemented optional parallelisation for all multi-query functions. Each multi-query function (with name ending in 'Dataframe' in **Table 1**) has an argument that allows the user to set a number of 'cores' to parallelise the query across. This can significantly reduce run-time, particularly for larger queries where the parallelisation overheads represent a small proportion of the overall memory requirements. For example, with a query size of 1000, the ensemblQueryLDwithSNPpairDataframe function running on a single core takes ~2.4 minutes, whereas using 10 cores speeds this up by 22 times, reducing execution time to ~0.11 minutes (System tested on: Ubuntu server 16.04 LTS with kernel version 4.4.0-210-generic, total RAM 251G).

Consistency of the data output format is an important feature of ensemblQueryR, making it amenable to R workflow integration. All functions (single- and multi-query) return a data frame object, including instances where the query returns a null result or an error. This consistent output format simplifies the process of writing custom code, allowing the user to incorporate ensemblQueryR functionality into bespoke workflows. It is important to note that when data cannot be retrieved, for example if a variant is not found in Ensembl databases, or the user input was invalid, console messages will alert the user of this and a data frame row containing 'NA' values will be returned for that query.

## Benchmarking

LDlinkR is an alternative R package that offers LD metric retrieval. As such, it was important to benchmark against this tool to demonstrate that our tool is an improvement in performance terms. Of the functions contained in the LDlinkR and ensemblQueryR packages, two functions are most comparable in terms of their functionality. Both the LDpair (from LDlinkR) and ensemblQueryLDwithSNPpair (from ensemblQueryR) aim to take a pair of rsIDs as input and output a table containing LD metrics for the query pair. As such, these were taken forward for benchmarking. To compare the performance of the two packages, computation speed and RAM usage at three query sizes representing a range of throughputs – 100, 1,000 and 10,000 queries – were assessed (**Figure 1**). For each function and query size combination, performance (speed and peak RAM usage) was tested 10 times to account for temporal fluctuations in processing speed and peak RAM usage, thus enabling accurate and reliable assessment of performance.

Firstly, comparing execution speed, we found that, on average (across the 10 tests), ensemblQueryLDwithSNPpair was 10.2 times faster in the 100 query test, taking an average of 0.208 minutes compared to 2.12 minutes for LDpair (**Figure 1b**). The 1,000 query test found that ensemblQueryLDwithSNPpair was, on average, 9.92 times faster than LDlinkR, taking an average of 1.97 minutes compared to 19.5 minutes for LDpair. Finally, in the 10,000 query test, LDpair was unable to produce a final results table in 7/10 tests and in these instances returning an error ('Bad Gateway (HTTP 502)'). In contrast, ensemblQueryLDwithSNPpair produced a final results table for 10/10 tests, demonstrating its reliability for high-throughput querying. Utilising the 3/10 successful tests of LDpair, we found that ensemblQueryLDwithSNPpair was, on average, 10.9 times faster than LDpair, taking an average of 18.5 minutes compared to 202 minutes (>3 hours). These speed improvements are likely due to server-side request rate limits which are higher for Ensembl, enabling fast concurrent or parallel requests.

Secondly, we compared peak RAM usage – the maximum RAM utilised at any one time during function execution – between ensemblQueryLDwithSNPpair and LDpair (**Figure 1a**). We found that across query sizes, ensemblQueryLDwithSNPpair had approximately a third (range: 20.8-49.6%) of the peak RAM usage of LDpair. These peak RAM usage improvements are likely due to a focus on within-function reduction of intermediate object storage and a reduction of the number of operations carried out within the ensemblQueryLDwithSNPpair function.

Taken together, we find that for analogous functions ensemblQueryLDwithSNPpair and LDpair, ensemblQueryR represents a performance improvement with respect to both speed (x10) and memory usage (1/3) over LDlinkR, underscoring the utility of our tool in the context of high-throughput workflows.

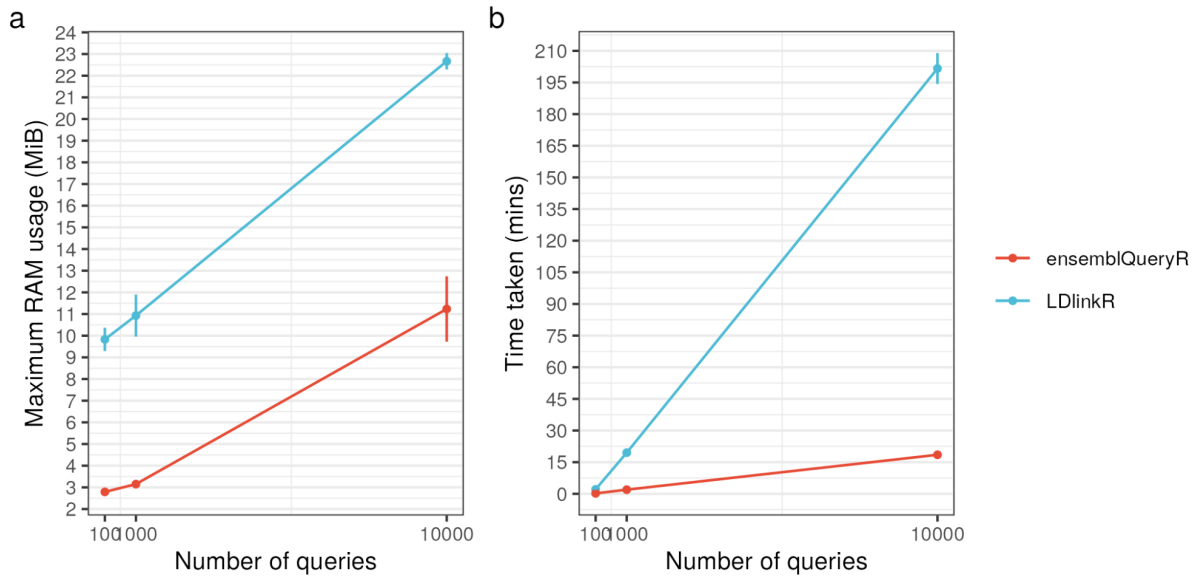

**Figure 1. Comparing performance metrics between analogous functions of the ensemblQueryR and LDlinkR R packages. a.** Plot to show, for query sizes of 100, 1,000 and 10,000, maximum RAM usage (mebibytes, MiB) during execution of ensemblQueryR's ensemblQueryLDwithSNPpair (red) and LDlinkR's LDpair (blue). **b.** Plot to show for query sizes of 100, 1,000 and 10,000, the execution time (minutes, mins) of ensemblQueryR's ensemblQueryLDwithSNPpair (red) and LDlinkR's LDpair (blue).

Usage

The following code provides an implementation example for ensemblQueryR. In this case, the target variant was taken from the OpenTargets homepage [14] with rsID rs4129267. To find out which variants, within a set genomic window size, are in LD ($R^2>0.8$ and $D'>0.8$) with the target variant, the ensemblQueryLDwithSNPWindow function was implemented as in **Figure 2**.

```r
# load package
library(ensemblQueryR)

# get list of available populations for which LD metrics can be
returned
ensemblQueryR::ensemblQueryGetPops()

# implement ensemblQueryLDwithSNPwindow to get variants in a window of
size 500kb centered on rs4129267, setting r2 and d.prime to constrain
the variants returned
ensemblQueryR::ensemblQueryLDwithSNPwindow(rsid="rs4129267",
  r2=0.8,
  d.prime=0.8,
  window.size=500,
  pop="1000GENOMES:phase_3:EUR")
```

**Figure 2. Working example of ensemblQueryR implementation.**

Methods

Docker and singularity images

In order to make this tool widely accessible to the research community, we have provided a Docker image (**Figure 3**). It is based on a Rocker [15] with R version 4.0.0 and Ubuntu version 20.04 LTS (Focal Fossa) pre-installed. To start a container using the 'ensemblqueryr' Docker image and launch an interactive R session within which the tool can be used, the following command can be executed on the command line:

```
docker pull ainefairbrotherbrowne/ensemblqueryr:1.0; \
docker run -t -d --name ensemblqueryr ainefairbrotherbrowne/ensemblqueryr:1.0; \
docker exec -i -t ensemblqueryr R
```

To mount a local volume that contains a file of variant IDs, for example, the following command can be used:

```
docker pull ainefairbrotherbrowne/ensemblqueryr:1.0; \
docker run -t -d --name ensemblqueryr ainefairbrotherbrowne/ensemblqueryr:1.0 -- path/to/volume; \
docker exec -i -t ensemblqueryr R
```

```dockerfile
# use rocker 4.0.0 and Ubuntu 20, as these resemble closely the development OS for ensemblQueryR
FROM rocker/r-ver:4.0.0

# system libraries of general use - informed by https://tinyurl.com/ypnkbrpb
RUN apt-get update && apt-get install -y \
    sudo \
    pandoc \
    pandoc-citeproc \
    pkg-config \
    libnlopt-dev \
    libcurl4-gnutls-dev \
    libcairo2-dev \
    libxt-dev \
    libgsl-dev \
    libssl-dev \
    libssh2-1-dev \
    libssl1.0 \
    libxml2-dev \
    openssl
RUN apt-get update && apt-get install -y \
    libmpfr-dev
RUN apt-get install libcurl4-openssl-dev -y

# install and load ensemblQueryR dependencies
RUN R -e "install.packages('remotes'); \
install.packages('httr'); \
install.packages('xml2'); \
install.packages('dplyr'); \
install.packages('jsonlite'); \
install.packages('purrr'); \
install.packages('tidyr'); \
install.packages('vroom'); \
install.packages('magrittr'); \
install.packages('parallel'); \
library(remotes); \
library(httr); \
library(xml2); \
library(dplyr); \
library(jsonlite); \
library(purrr); \
library(tidyr); \
library(vroom); \
library(magrittr); \
library(parallel)"

# install and load ensemblQueryR (install from clone)
RUN apt-get install -y git
RUN git clone https://github.com/ainefairbrother/ensemblQueryR.git ./ensemblQueryR
RUN R -e "install.packages('devtools'); devtools::install('./ensemblQueryR'); library(ensemblQueryR)"

# clean up after installer
RUN rm -rf /var/lib/apt/lists/*
```

**Figure 3. Dockerfile containing the setup for the ensemblqueryr Docker image.**

Additionally, for use-cases wherein the user does not have sufficient permissions to use Docker, such as on HPC clusters, a singularity image can be used instead. To generate a singularity image, we converted the ensemblqueryr docker image [11] into a singularity image using the Docker-based tool docker2singularity (Version 2.6) [16]. This image can be found on the

sylabs.io singularity repository [12]. The singularity image can be pulled and executed, launching a command line R session as follows:

```
singularity pull --arch amd64 
library://ainefairbrother/ensemblqueryr/ensemblqueryr:sha256.e387ea11ae4eaea8f9
4d81c625c2c1d5a22dd351858ebcd03910a7736d76ca30; \
singularity exec 
ensemblqueryr_sha256.e387ea11ae4eaea8f94d81c625c2c1d5a22dd351858ebcd03910a7736d
76ca30.sif R
```

Benchmarking

To benchmark our package against LDlinkR, we first selected functions that performed similar tasks. The most comparable functions in this context were LDlinkR's 'LDpair' and ensemblQueryR's 'ensemblQueryLDwithSNPpair'. Both are configured for single queries and take a pair of variant IDs as input, retrieving the corresponding LD metrics through an API query. To run multiple queries using these single-query functions, both functions were applied over variant identifier (ID) vectors using the base R lapply (Version 4.0.5) [13].

To compare the utility of these functions for LD metric querying in the context of high-throughput workflows, we measured the peak RAM usage and execution speed using the R package peakRAM (version 1.0.2) [17] 10 times for each function and at three query sizes (100, 1000 and 10000). The standard deviation (sd) of the performance metric for each function at each timepoint across the 10 iterations was calculated. Mean performance metrics were plotted with error bars to show the standard deviation from the mean for peak RAM usage (**Figure 1a**) and time taken in minutes (**Figure 1b**) for the three query sizes.

Limitations

It is important to note that there are some limitations to this R package. One is that although this package enables high-throughput querying of the Ensembl API, there is an inherent limit to the number of queries that can be submitted arising from the API query limit which is set at 54,000 requests per hour [5,6]. On the Ubuntu system used to develop ensemblQueryR (Ubuntu server 16.04 LTS with kernel version 4.4.0-210-generic, total RAM 251G), 54,000 API requests via ensemblQueryLDwithSNPpairDataframe took ~1.93 hours on a single core, making the per-hour request rate 27,867. As such, even query sizes of 54,000 can be run unparallelised and are unlikely to exceed the Ensembl API hourly rate. However, this request limit must be considered by users when applying parallelisation to large queries. The second limitation is that the parallelisation library used to enable the multi-core functionality is the R package 'parallel' (version 3.6.2) [18] which works on OSIX systems (Mac, Linux and other Unix-based operating systems) but does not work on Windows.

## Scope for future development

At present, this package provides wrappers for the Ensembl API endpoints that retrieve LD data. However, the Ensembl API offers ~109 other endpoints [5], all of which have the potential to be wrapped into R functions and included in this package. As such, there is scope for the usefulness of this package beyond LD metrics and further development will expand its utility to R users across an array of bioscience disciplines.

## Availability of source code and requirements

- Project name: ensemblQueryR
- Project home page: https://github.com/ainefairbrother/ensemblQueryR
- Operating system(s): Platform independent
- Programming language: R
- Licence: MIT

## Data availability

Not applicable

## List of abbreviations

LD: Linkage Disequilibrium
RAM: Random Access Memory
REST: Representational State Transfer
API: Application Programming Interface
GWAS: Genome-Wide Association Study
eQTL: Expression Quantitative Trait Loci
rsID: Reference SNP cluster IDs
sd: standard deviation

## Declarations

Not applicable

### Ethical approval

Not applicable

# Competing interests

R.H.R. is currently employed by CoSyne Therapeutics (Lead Bioinformatician). All work performed for this publication was performed in her own time, and not as a part of her duties as an employee.

# Funding

A.F.-B. was supported through the award of a Biotechnology and Biological Sciences Research Council (BBSRC UK) London Interdisciplinary Doctoral Fellowship. S.G.R., and M.R. were supported through the award of a Tenure Track Clinician Scientist Fellowship to M.R. (MR/N008324/1). A.H. was funded by a BBSRC award (BB/R006075/1).

# Author contributions

- Conceptualisation: AFB, RHR
- Software: AFB
- Docker image: AFB, SGR
- Benchmarking: AFB
- Writing - original draft: AFB
- Writing - review and editing: AFB, SGR, AH, MR, RHR
- Supervision: MR, AH

# Acknowledgements

N/A

# Endnotes

N/A